\documentclass{elsart}
\usepackage{epsfig}
\date{22 June 2005}

\begin{document}

\begin{frontmatter}

\title{Study of  a self quenching streamer  mode in pure  gases of DME
and isobutane}
\thanks[correspondence]{Corresponding author.  \\ 
 E-mail address: davydov@triumf.ca (Yuri Davydov)}
\author[a,b]{Yu.I.~Davydov\thanksref{correspondence}},
\author[b]{R.~Openshaw}  
\address[a]{RRC "Kurchatov Institute", 
Kurchatov sq. 1, Moscow, 123182, Russia} 
\address[b]{TRIUMF, 4004 Wesbrook Mall, Vancouver, BC, Canada  V6T 2A3 }

\begin{abstract}
A self  quenching streamer  (SQS), or limited  streamer mode  has been
studied    in    single     wire    chambers    with    cross-sections
12$\cdot$12~mm$^{2}$    and    wire     diameters    15,    25    and
50~$\mu$m. Chambers were filled  with either pure dimethyl ether (DME)
or isobutane gases and  irradiated with $^{148}$Gd alpha and $^{55}$Fe
x-ray  sources.  Clear transitions  from a  proportional to  100\% SQS
mode  were  observed on  all  three  diameter  wires with  both  gases
irradiated  with  alpha  particles.   Double  SQS  discharges  due  to
inclined  tracks  observed in  DME  gas  allowed  an estimation  of  a
streamer  size  along the  wire  of less  than  1~mm.  The second  SQS
discharge appears less  than 1~mm from the first  within about 500~ns.
Charge  spectra obtained  with  DME irradiated  with $^{55}$Fe  x-rays
might also be  interpreted as a transition to a  SQS mode, although no
direct evidence of that was seen in the observed pulse shapes.


\end{abstract}


\end{frontmatter}

\section{Introduction}

\indent Since the first  observation of self quenching streamer (SQS),
or   limited  streamer   signals~\cite{Charpak,Brehin},   most  models
attempting to  explain this discharge  have employed the  mechanism of
photon ionization  of the gas~\cite{Alekseev_1,Atac,Zhang}.   In fact,
most  studies of  a  SQS mode  have  been done  with  noble gas  based
mixtures. The high probability of gas ionization by photons emitted by
the excited atoms  makes it possible to produce SQS  signals due to an
initial  ionization   by  x-rays  or   beta-particles.   However,  the
observation of SQS discharges  in pure quenching gases of hydrocarbons
irradiated  with  alpha particles~\cite{Iwatani,Koori_1}  demonstrated
the more complicated nature of this type of discharge. \\
\indent This work presents a study of the transition from proportional
to SQS mode  in single wire chambers filled  with either pure dimethyl
ether   (DME)  or   \mbox{iso-C$_4$H$_{10}$}  gases   irradiated  with
$^{148}$Gd  alpha  particles  or   $^{55}$Fe  x-rays.   Both  DME  and
\mbox{iso-C$_4$H$_{10}$} are excellent  quenchers and should provide a
transition  from  a proportional  to  SQS  mode  without influence  of
photons.  This  could provide data  for a better understanding  of the
nature of this discharge. 

\section{Experimental setup}

\indent Single wire chambers with  15, 25 and 50~$\mu$m diameter wires
were  used  to  carry  out   the  tests.   The  chambers  have  square
12$\cdot$12~mm$^{2}$ cross  sections and  were made of  aluminum alloy
with  6.35~$\mu$m thick aluminized  mylar windows  on two  sides. Wire
lengths are approximately 20~cm. \\
\indent  Chambers  were  irradiated  with 3.183~MeV  $^{148}$Gd  alpha
particles and 5.9~keV $^{55}$Fe x-rays.  During most test measurements
sources were placed directly over the wires on a 1.6~mm thick aluminum
plate with  1.2~mm diameter collimating  hole.  The distance  from the
source to  the cathode foil  is about 2.3~mm.  Under  these conditions
alpha  particles entering normal  to the  chamber lose  about 1.65~MeV
energy  in the  chamber gas  volume.   Their track  lengths are  about
2.35~mm in  chambers filled with  \mbox{iso-C$_4$H$_{10}$}, and 3.1~mm
with DME.  Alpha particles produce about 7$\cdot$10$^{4}$ ion-electron
pairs, while the  x-ray particles produce about 250  pairs.  The alpha
particle track  ranges and energy  losses were estimated with  the ion
range-energy  code SRIM~\cite{SRIM}.  Signals  from the  chambers were
self triggered and fed into a  LeCroy 2249W ADC.  The ADC gate signals
had  a  duration of  2.5~$\mu$s  for  alpha  particle irradiation  and
1.5~$\mu$s for x-rays. 

\section{Results and discussion}

\begin{figure}[t]
\centering
\begin{minipage}[c]{0.8\textwidth}
\vspace{5mm}
\begin{center}
\epsfig{file=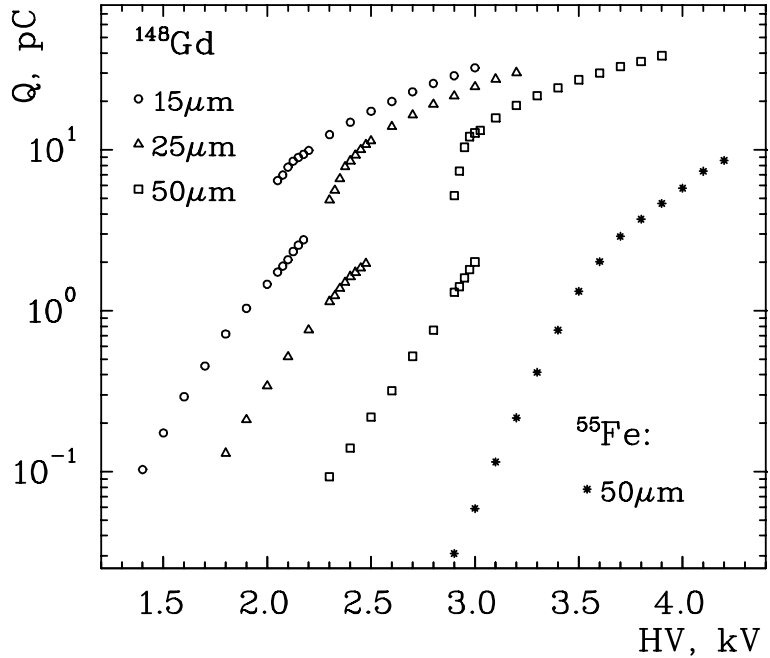,width=10cm,angle=0}
\end{center}
\caption{Measured  charge as  a function  of  high voltage  in pure
\mbox{iso-C$_4$H$_{10}$}  filled chambers  with 15,  25  and 50~$\mu$m
diameter anode  wires irradiated  with $^{148}$Gd alpha  particles and
50~$\mu$m wire irradiated with $^{55}$Fe x-rays.}
\label{fig1}
\end{minipage}\hfill
\end{figure}

\begin{figure}[t]   
\centering
\begin{minipage}[c]{0.8\textwidth}
\vspace{5mm}
\begin{center}   
\epsfig{file=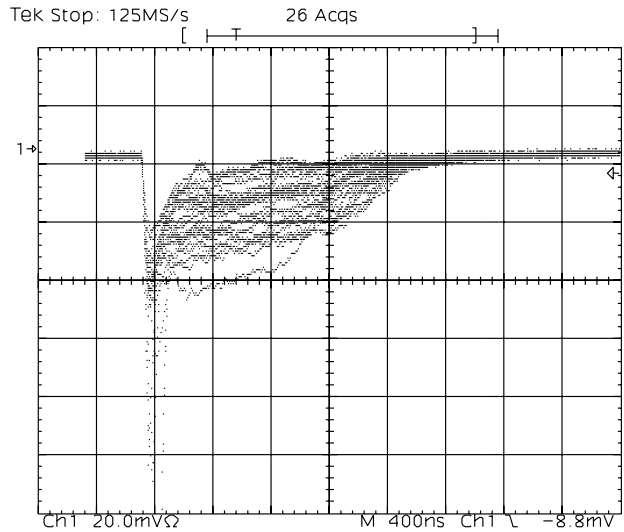,width=10cm,angle=0}
\end{center}
\caption{Amplified signals  on the  DME filled chamber  with 15~$\mu$m
diameter  wire  at  2050  V,  irradiated with  alpha  particles.  Wide
proportional and  narrow SQS signals with high  amplitude are clearely
seen on the graph.}
\label{fig2}
\end{minipage}\hfill
\end{figure}

\begin{figure}[t]
\centering
\begin{minipage}[c]{0.8\textwidth}
\vspace{5mm}
\begin{center}
\epsfig{file=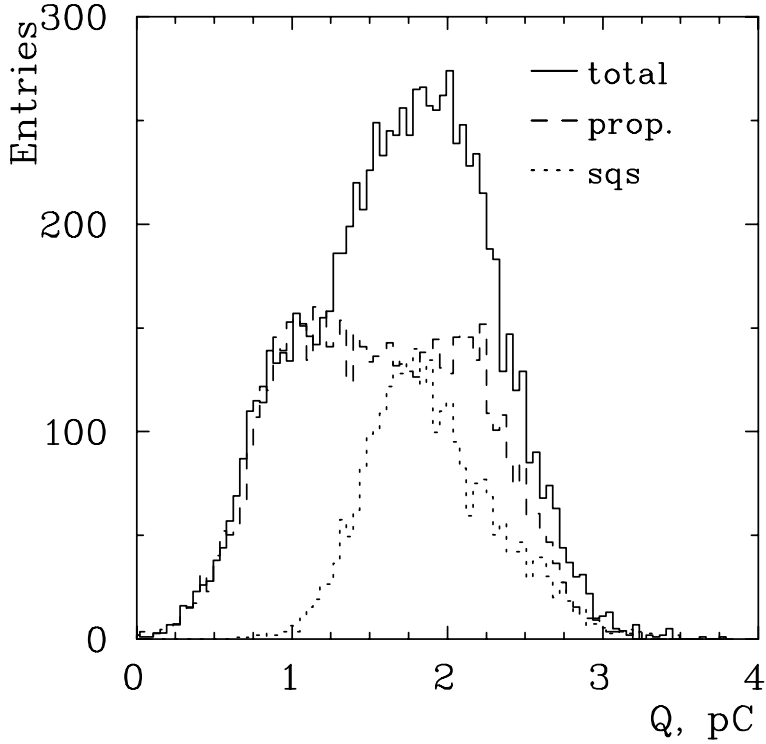,width=10cm,angle=0}
\end{center}
\caption{Charge  spectra  on the  DME  filled  chamber with  15~$\mu$m
diameter wire at 2050 V  irradiated with alpha particles. Total charge
and contributions of proportional and  SQS signals to the total charge
spectrum are shown.}
\label{fig3}
\end{minipage}\hfil
\end{figure}

\begin{figure}   
\centering
\begin{minipage}[c]{0.8\textwidth}
\vspace{5mm}
\begin{center}
\epsfig{file=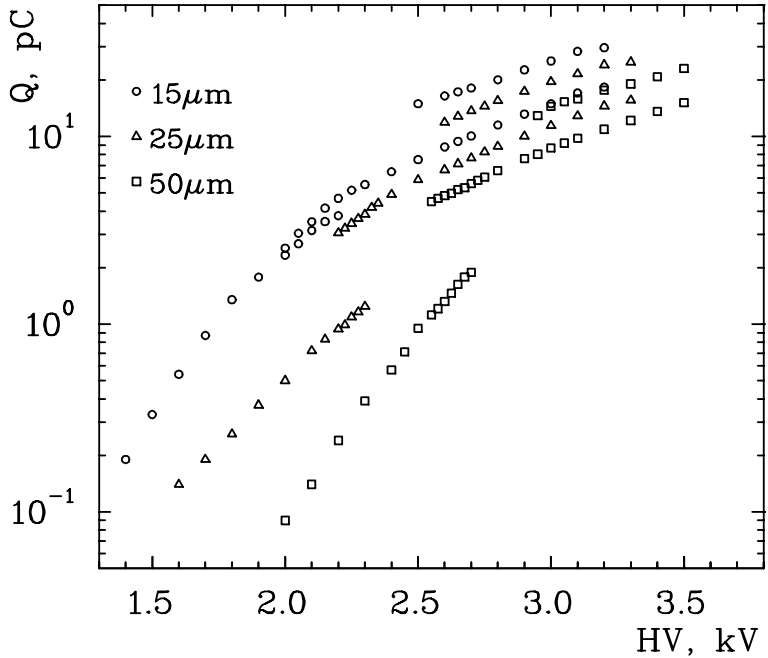,width=10cm,angle=0}
\end{center}
\caption{Measured charge as a function of applied high voltage for DME
filled chambers  with 15, 25  and 50~$\mu$m diameter  wires irradiated
with alpha particles.}
\label{fig4}
\end{minipage}\hfil
\end{figure}

\begin{figure}
\centering
\begin{minipage}[c]{0.8\textwidth}
\vspace{5mm}
\begin{center}
\epsfig{file=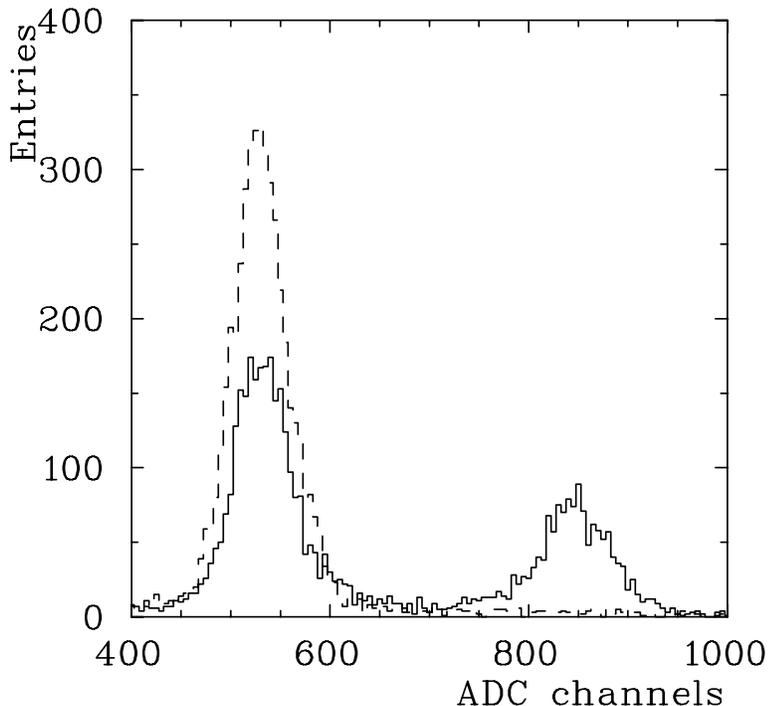,width=10cm,angle=0}
\end{center}
\caption{Charge  spectra from  the DME  filled chamber  with 15~$\mu$m
diameter  wire  at  2900~V.  The  chamber was  irradiated  with  alpha
particles  with a  0.5$\cdot$5~mm  slot  collimator. The  long  side of  the
collimator  was set  perpendicular  (dashed line)  or parallel  (solid
line) to the wire.  There is one SQS peak in the  first case while the
latter has two peaks.}
\label{fig5}
\end{minipage}\hfil
\end{figure}

\begin{figure}
\centering
\begin{minipage}[c]{0.8\textwidth}
\vspace{5mm}
\begin{center}
\epsfig{file=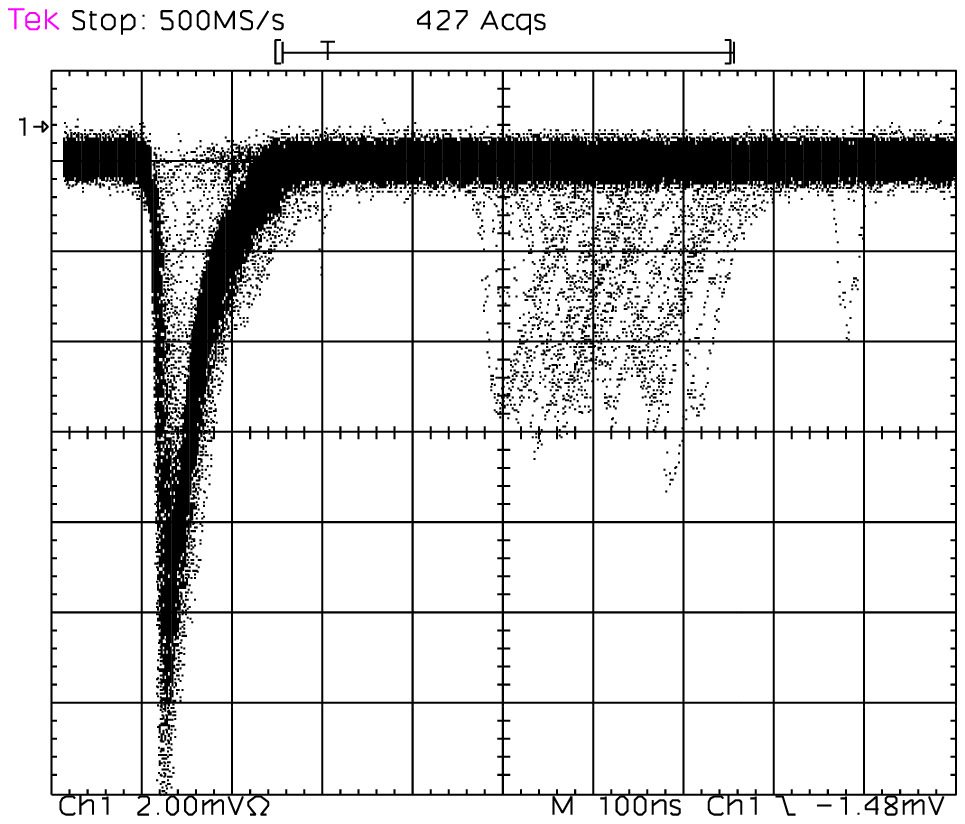,width=10cm,angle=0}
\end{center}
\caption{Signals  on  the  50~$\mu$m  diameter  wire  at  3100~V  with
50~$\Omega$ load. Chamber is filled  with DME.  Second SQS signals are
delayed by a few hundred nanoseconds and have smaller amplitudes.}
\label{fig6}
\end{minipage}\hfil
\end{figure}

\begin{figure}
\centering
\begin{minipage}[c]{0.8\textwidth}
\vspace{5mm}
\begin{center}
\epsfig{file=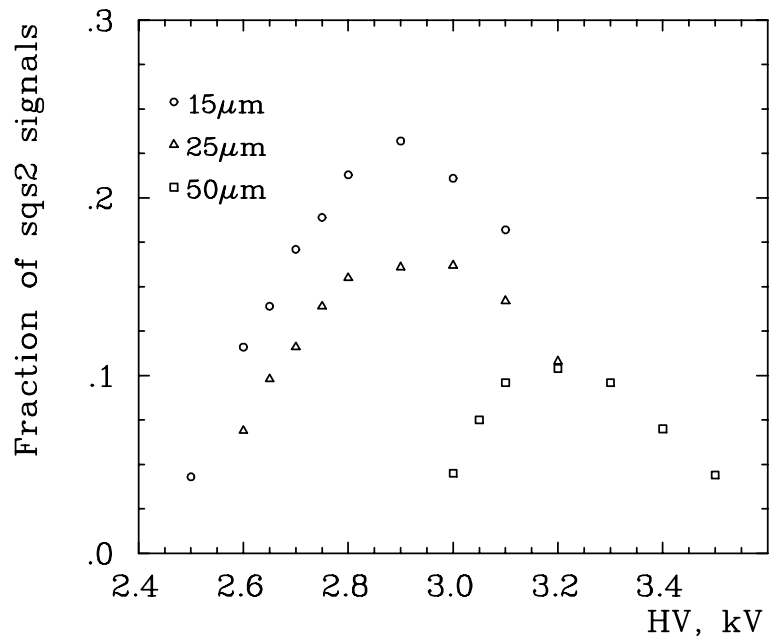,width=10cm,angle=0}
\end{center}
\caption{Fraction of events in the second SQS peak (sqs2) as function of
a high voltage for  15, 25 and 50~$\mu$m diameter  wires irradiated with
alpha particles. Chambers are filled with DME.}
\label{fig7}
\end{minipage}\hfil
\end{figure}

\begin{figure}
\centering
\begin{minipage}[c]{0.8\textwidth}
\vspace{5mm}
\begin{center}
\epsfig{file=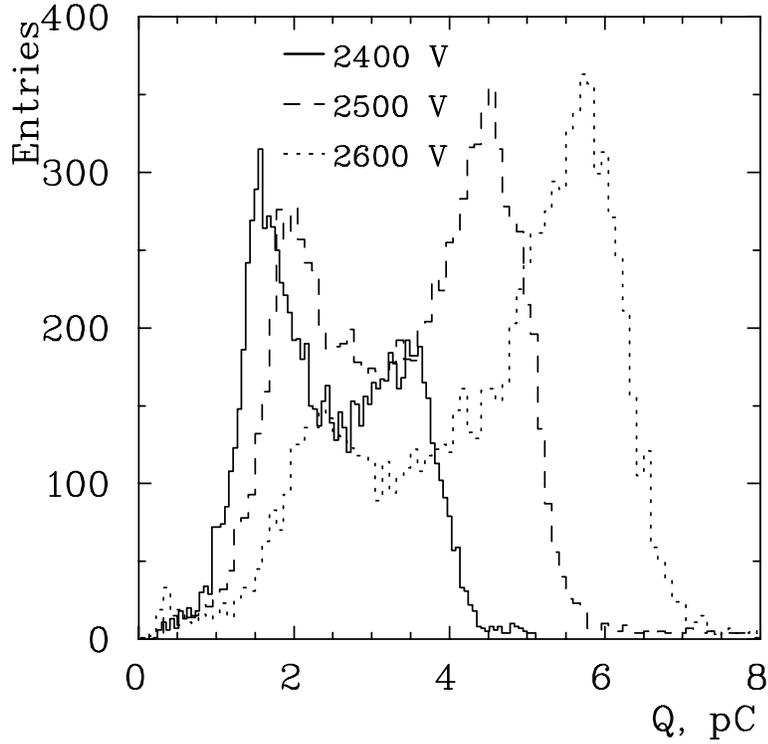,width=10cm,angle=0}
\end{center}
\caption{Charge  spectra of  the  DME filled  chamber with  15~$\mu$m
diameter wire  at 2400~V, 2500~V and 2600~V  irradiated with $^{55}$Fe
x-rays.}
\label{fig8}
\end{minipage}\hfil
\end{figure}

\begin{figure}
\centering
\begin{minipage}[c]{0.8\textwidth}
\vspace{5mm}
\begin{center}
\epsfig{file=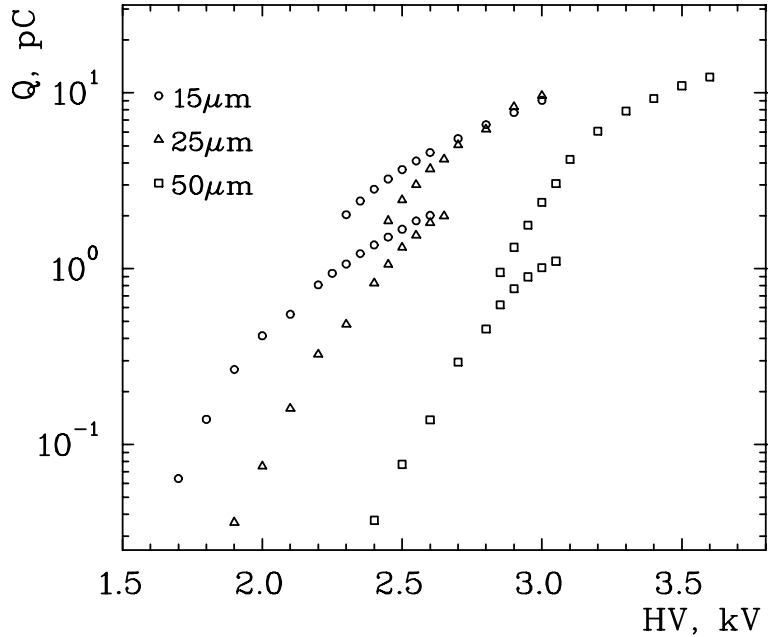,width=10cm,angle=0}
\end{center}
\caption{Collected charge as a function of applied high voltage on 15,
25   and   50~$\mu$m   diameter   wires  irradiated   with   $^{55}$Fe
x-rays. Chambers are filled with DME.}
\label{fig9}
\end{minipage}\hfil
\end{figure}

\indent Charge spectra obtained on  all three wire diameters with pure
\mbox{iso-C$_4$H$_{10}$} irradiated with alpha particles, became wider
in  proportional mode  upon increase  of high  voltage. This  could be
explained by  the wide angle distribution of  alpha particles entering
the  chamber volume.  Further  high voltage  increase resulted  in the
appearance  of   SQS  signals  and   then  a  100\%   transition  from
proportional to  SQS mode on  all three tested  wires.  On all  3 wire
diameters, the peaks of the SQS signals on the charge spectra are well
separated  from the  proportional signals.   However, all  SQS spectra
have tails in  the lower portion of charge spectra,  down to the level
of the proportional signals.   Fig.\ref{fig1} shows measured charge as
a function  of applied high voltage  on 15, 25  and 50~$\mu$m diameter
wires irradiated  with the alpha source.  Mean  values of proportional
and SQS spectra  are plotted. The few points  in the transition region
and above the mentioned tails in the SQS charge spectra could indicate
that  SQS  signals  had  not  been fully  developed  at  this  voltage
range. \\
\indent Charge spectra  from the chamber with a  50~$\mu$m wire filled
with pure \mbox{iso-C$_4$H$_{10}$} and irradiated with $^{55}$Fe x-ray
source do  not show any  transition to SQS mode  (see fig.\ref{fig1}).
Our measurements for the chamber  with 50~$\mu$m wire filled with pure
\mbox{iso-C$_4$H$_{10}$} are  consistent both for  alpha particles and
x-rays with \cite{Koori_1}, where a single wire chamber with a similar
geometry    had   been   employed    (50~$\mu$m   wire,    cell   with
10.5$\cdot$10.5~mm$^{2}$ cross-section). \\
\indent Similar to the pure \mbox{iso-C$_4$H$_{10}$}, in pure DME with
alpha particle irradiation of all  three tested wires, the increase of
high  voltage  on the  chambers  in  the  proportional mode  at  first
resulted in  a widening  of the charge  spectra.  Further  increase of
high voltage produced a clear transition from the limited proportional
to SQS mode on a 50~$\mu$m wire, resulting in the appearance of a well
separated second peak on the  charge spectrum. For a 25~$\mu$m wire, a
second peak overlapping the upper portion of the main peak appeared in
the charge spectrum  as the chamber high voltage  was increased.  Upon
further increase in high voltage, all signals moved up into the second
peak.   Observation  of  the   pulses  with  an  oscilloscope  clearly
demonstrated that these  were SQS signals. They had  shorter width and
much higher amplitude than proportional signals. \\
\indent Charge  spectra on a 15~$\mu$m  diameter wire do  not show any
additional peaks  upon increase of  high voltage.  However,  clear SQS
signals were observed on an oscilloscope at a high voltage starting at
about  2000  V.   Fig.\ref{fig2}  presents amplified  signals  from  a
15~$\mu$m  diameter wire  at  U=2050 V.   The  oscilloscope scale  was
configured  to  provide details  of  the  proportional signals.   Long
proportional  and short SQS  signals with  high amplitude  are clearly
seen on the graph. \\
\indent  In  order  to  derive  the  relative  fractions  of  SQS  and
proportional signals,  charge spectra  have been taken  with different
thresholds in the transition regions for 15~$\mu$m and 25~$\mu$m wires
(low threshold for the total distribution, high threshold for SQS, and
differential  threshold  for  proportional  signals).   Fig.\ref{fig3}
presents the  total charge spectrum  on the 15~$\mu$m wire  at 2050~V,
and  shows the  separate  contributions of  the  SQS and  proportional
signals to  the total  spectrum.  The charge  spectrum of  SQS signals
overlaps the wider charge spectrum of proportional signals. \\
\indent Further increase in high voltage resulted in the appearance of
a second SQS peak on the charge spectra of all wires.  This second SQS
peak comprises  a small fraction of  all events and  is well separated
from the first SQS peak for all three wires. \\
\indent  Fig.\ref{fig4} presents measured  charge versus  applied high
voltage for all  three wire diameters. Mean values  of measured charge
spectra for both proportional  and SQS signals are plotted.  25~$\mu$m
and 50~$\mu$m  diameter wires  have big jumps  in the  measured charge
when moving  from the  proportional to SQS  mode, while  the 15~$\mu$m
wire has almost no jump.  The  jumps are almost the same magnitude for
all three  wires when  moving from  the first SQS  peak to  the second
one. \\
\indent Further tests revealed that the  second SQS peak is due to the
development  of a  second SQS  discharge  along the  wire, from  alpha
particles entering the chamber volume at big angles, with tracks which
are nearly  co-planar with  the wire.  Fig.\ref{fig5}  presents charge
spectra taken from the chamber  with 15~$\mu$m diameter wire at 2900~V
with  a  0.5$\cdot$5~mm$^{2}$  slot  collimator.  In  first  case  the
collimator  slot   was  placed  perpendicular  to   the  wire  (dashed
line). The  second case presents  data for the collimator  slot placed
parallel  to the  wire (solid  line).  The  $^{148}$Gd  alpha particle
source was  set at  the same  height over the  chamber foil  as during
tests with  an aluminum  plate hole collimator.   The second  SQS peak
appears  only  in  the case  of  a  slot  collimator parallel  to  the
wire. These second SQS signals  are delayed relative to the first ones
by a few hundred nanoseconds and have a smaller amplitude, as shown on
fig.\ref{fig6} for 50~$\mu$m diameter  wire at 3100~V with 50~$\Omega$
load.  Note that  the shortest drift time from the  cathode to wire at
this  applied voltage,  estimated  using Garfield~\cite{Garfield},  is
longer  than  800~ns.    Also,  delayed  proportional  signals  appear
initially on the wires at lower voltages before they transfer into the
second SQS discharge as the voltage is increased. \\
\indent The source placed over the collimator plate with a 1.2~mm hole
emits alpha  particles with  angles up to  37$^\circ$ relative  to the
normal.   The alpha  particles entering  the chamber  at  this maximum
angle  have a  track length  inside  the chamber  about 2.15~mm.   The
maximum  projection of that  track on  the wire  occurs when  they are
co-planar, and is equal to about 1.3~mm. Two streamers developing from
these inclined  tracks indicate that  streamer size along the  wire is
less  than 1~mm.  Moreover,  this demonstrates  that the  wire becomes
sensitive to  another discharge, at  less than 1~mm distance  from the
initial   streamer,  within   about  500~ns,   as  can   be   seen  on
fig.\ref{fig6}. The dead  zone values, defined as the  product of dead
length and  dead time, obtained by  other authors for  noble gas based
mixtures~\cite{Alekseev_2,Koori_2,Nohtomi},  are  orders of  magnitude
bigger than that observed for DME.  This suggest that photons from the
excited noble gas  atoms play a major role  in increasing the streamer
size along the  wire, increasing both dead length  and especially dead
time. \\
\indent Fig.\ref{fig7}  presents the fraction of events  in the second
SQS peak (sqs2) for three wire diameters as a function of applied high
voltage.  For all three wire  diameters, the fraction of events in the
second  peak reaches a  maximum value  at some  high voltage  and then
decreases with further increase  of high voltage.  Two processes cause
the drop in  the fraction of events in the  second peak.  Firstly, the
dead zone  increases as the first streamer  size increases.  Secondly,
in unsaturated  DME gas, the drift  time spread of  electrons from the
track decreases  with increasing high voltage,  and becomes comparable
to the dead time. The thinner  wire has the highest fraction of events
in the  second peak  due to  having the longest  drift time  spread at
these applied voltages. \\
\indent Many  other authors  have observed multi  streamer discharges.
To our knowledge, these multistreamer discharges were observed only in
noble gas based  mixtures.  In those mixtures the  new streamers occur
due to photons emitted by the excited noble gas atoms.  In the case of
a gas  photoionization, the following  streamers are indistinguishable
in  time from  the initial  ones~\cite{Alekseev_2}.   When afterpulses
appear  due to  a photoelectric  effect  on the  cathode surface,  the
following streamers  are separated by  electron's drift time  from the
cathode to anode wire~\cite{Battistoni,deLima,Avanzini}. \\
\indent In our case with DME,  the second SQS discharge along the wire
occurs due  soley to  the initial ionization  from the  alpha particle
tracks.    The    absence   of   second   SQS    discharges   with   a
0.5$\cdot$5~mm$^{2}$ slot collimator  placed perpendicular to the wire
(see fig.~\ref{fig5}) confirms that  photoionization does not play any
role  in  the  formation  of  SQS  discharge  in  pure  DME  at  these
voltages. \\
\indent A similar behaviour  was observed for chambers with 15~$\mu$m,
25~$\mu$m and 50~$\mu$m diameter  wires filled with DME and irradiated
with $^{55}$Fe x-rays. Increasing the chamber high voltage resulted at
first in the  appearance of a tail on the upper  portion of the charge
spectra, and then of a second peak overlapping that tail. Signals from
the first peak moved up into  the second peak upon further increase in
high  voltage.   Fig.\ref{fig8} shows  these  features  of the  charge
spectra on  the 15~$\mu$m  diameter wire  at 2400 V,  2500 V  and 2600
V. \\
\indent   Fig.\ref{fig9}  shows   collected   charge  from   $^{55}$Fe
irradiaton as a function of high voltage for all three wire diameters.
The  data points  represent the  most  probable values  of the  charge
spectra.  One can see that  the charge characteristics look similar to
the  ones  obtained by  irradiation  with  an  alpha particle  source.
However,  the curves will  have smooth  shapes if  one plots  the mean
values of the total  charge spectra.  Observation with an oscilloscope
did  not  show  clear  differences  between the  signals  in  the  two
peaks. \\
\indent  The observed  charge  spectra on  all  wires irradiated  with
$^{55}$Fe x-rays could be due  to the properties of DME.  The widening
of  charge  spectra of  proportional  signals  upon  increase of  high
voltage on all  three wires irradiated with alpha  particles, as shown
on  fig.\ref{fig3} for  15~$\mu$m wire,  supports this  view.   On the
other hand it  could be a real transition to SQS  mode. The signals in
the  two  peaks  are  not distinguished  by  oscilloscope  observation
because the drift time spread  of electrons produced in DME by 5.9~keV
x-ray ionization is  about 50~ns, which is comparable  to the duration
of the typical SQS signals. 

\section{Conclusion}

\indent We have  studied a transition from a  proportional to SQS mode
in  single wire  chambers with  wire diameters  15, 25  and 50~$\mu$m,
filled with either pure DME or iso-C$_4$H$_{10}$ gases.  Chambers were
irradiated with  $^{148}$Gd alpha  or $^{55}$Fe x-ray  particles. \\
\indent  All   three  wires   demonstrate  a  100\%   transition  from
proportional to SQS mode in  both gases for the ionization produced by
alpha particles.  All  wires show a second SQS peak  with DME gas with
further increase of  high voltage.  It was shown  that the second peak
is  due to  the inclined  tracks allowing  a second  SQS  discharge to
develop  along the  wire.   No  second SQS  peaks  were observed  with
iso-C$_4$H$_{10}$ gas irradiated with the same alpha source. \\
\indent Double streamer discharges indicate  that the wire in pure DME
irradiated by  alpha particles has  much smaller dead length  and dead
time compared to those of  noble gas based mixtures. The streamer size
along the wire  is less than 1~mm. The second  SQS discharge occurs at
less than 1~mm distance within approximately  500~ns. \\
\indent The tests  with a slot collimator suggest  that in the applied
voltage ranges used, photons do not  play any role in SQS formation in
pure DME. \\
\indent Charge distributions observed on all three diameter wires with
DME filling  irradiated by $^{55}$Fe x-rays are  not fully understood.
Increasing high voltage resulted first  in the appearance of a tail on
the upper  portion of the charge  spectrum, followed by  a second peak
overlapping that tail.  All signals moved up into the second peak upon
further  increase of  high voltage.   This could  be interpreted  as a
transition from a proportional to SQS mode, however the observation of
signals on  the oscilloscope did  not provide clear evidence  for this
premise. \\
\indent The observed difference in charge characteristics of chambers,
filled with \mbox{iso-C$_4$H$_{10}$} and DME and irradiated with alpha
particles,  are  most  likely  due  to the  different  electron  drift
velocities in  the two gases and  as a result,  different space charge
development dynamics.

\end{document}